\documentclass[aps,prl,reprint,twocolumn,superscriptaddress,floatfix,nofootinbib,longbibliography]{revtex4-1}
\usepackage{amsthm}
\usepackage{amsmath,amssymb,color,comment,physics}
\usepackage[makeroom]{cancel}
\usepackage[caption=false]{subfig}
\usepackage{mathrsfs}
\usepackage{graphicx}
\usepackage{subfig}
\usepackage[countmax]{subfloat}
\usepackage[english]{babel}
\usepackage{dsfont}
\usepackage[bookmarks=true,colorlinks,linkcolor=OrangeRed,urlcolor=NavyBlue,citecolor=RoyalBlue]{hyperref}
\usepackage[dvipsnames]{xcolor}
\usepackage{braket}

\definecolor{mygold}{rgb}{0.93,0.59,0.13}
\definecolor{mypurple}{rgb}{0.49,0.18,0.56}

 \newcommand{\eq}[1]{\begin{align}#1\end{align}}

\usepackage{ulem}
\definecolor{philipp}{rgb}{1,.4,.3}

\newtheorem{theorem}{Theorem}
\newtheorem{lemma}[theorem]{Lemma}

\definecolor{mygreen}{rgb}{0.25,0.5,0.25}

\begin{document}

\preprint{APS/123-QED}

\title{Probing dynamical criticality near quantum phase transitions}

\author{Ceren B.~Da\u{g}}
\email{ceren.dag@cfa.harvard.edu}
\thanks{C.D.~and Y.W.~contributed equally to this work.}
\affiliation{ITAMP, Harvard-Smithsonian Center for Astrophysics, Cambridge, Massachusetts, 02138, USA}
\affiliation{Department of Physics, Harvard University, 17 Oxford Street Cambridge, MA 02138, USA}

\author{Yidan Wang}
\thanks{C.D.~and Y.W.~contributed equally to this work.}
\affiliation{Department of Physics, Harvard University, 17 Oxford Street Cambridge, MA 02138, USA}

\author{Philipp Uhrich}
\affiliation{INO-CNR BEC Center and Department of Physics, University of Trento, Via Sommarive 14, I-38123 Trento, Italy}

\author{Xuesen Na}
\affiliation{Simons Laufer Mathematical Sciences Institute, 17 Gauss Way, Berkeley, CA 94720, USA.}

\author{Jad C.~Halimeh}
\affiliation{INO-CNR BEC Center and Department of Physics, University of Trento, Via Sommarive 14, I-38123 Trento, Italy}

%\date{\today}

\begin{abstract}
We reveal a prethermal temporal regime upon suddenly quenching to the vicinity of a quantum phase transition in the time evolution of 1D spin chains. The prethermal regime is analytically found to be self-similar, and its duration is governed by the ground-state energy gap. Based on analytical insights and numerical evidence, we show that this \textit{critically prethermal regime} universally exists independently of the location of the probe site, the presence of weak interactions, or the initial state. Moreover, the resulting prethermal dynamics leads to an out-of-equilibrium scaling function of the order parameter in the vicinity of the transition.
\end{abstract}

\maketitle

Out-of-equilibrium quantum many-body physics has recently been at the forefront of theoretical and experimental investigations in condensed matter physics \cite{Eisert_2015} due to recent impressive progress in the control and precision achieved in quantum synthetic matter \cite{Greiner_2002,2009Natur.462...74B,Cheneau_2012,islam2015measuring,Kaufman_2016,Zhang_2017,G_rttner_2017}. Not only have concepts from equilibrium physics been extended to the out-of-equilibrium realm such as with dynamical phase transitions \cite{Mori_2018,PhysRevLett.110.135704,PhysRevLett.110.136404,Jurcevic2017,Flaeschner2018} and dynamical scaling laws \cite{PhysRevLett.110.136404,PhysRevB.88.201110,PhysRevLett.115.245301,PhysRevB.91.220302,PhysRevA.100.031601,2020arXiv200412287D,2020arXiv200506481T}, but there have also been concerted efforts to probe equilibrium quantum critical points (QCPs) and universal scaling laws through quench dynamics \cite{PhysRevLett.98.180601,PhysRevB.88.201110,PhysRevB.91.220302,PhysRevE.96.022110,PhysRevLett.121.016801,PhysRevA.100.013622,PhysRevLett.123.115701,PhysRevB.101.245148,2020arXiv200412287D,PhysRevX.11.031062} or with infinite-temperature initial states \cite{PhysRevB.97.235134,PhysRevB.101.104415,https://doi.org/10.1002/andp.201900270}. Such techniques obviate the need for undertaking the usually difficult task of cooling the system into its ground state over a range of its microscopic parameters in order to construct its equilibrium phase diagram. The underlying concept behind these works is of the Landau paradigm \cite{Mori_2018}, i.e., it is based on nonanalytic behavior in the long-time dynamics of a local order parameter. This indicates that, in principle, such nonanalytic behavior may be used to extract equilibrium criticality that manifests itself dynamically. 

It is well known that relaxation times of order parameters (OP) diverge at QCP after adiabatic quenches \cite{doi:10.1080/00018732.2010.514702,RevModPhys.83.863}. Such divergent behavior of the OP is a signature of the nonanalyticity at the QCP and is often referred as \textit{critical slowing down}. Although the current focus of the literature is utilizing sudden quenches to probe the QCP, how the relaxation time of the OP behaves around the QCP after a sudden quench has not been sufficiently explored \cite{doi:10.1080/00018732.2010.514702,PhysRevLett.103.056403,PhysRevLett.102.130603,PhysRevB.81.115131,PhysRevB.91.220302,2020arXiv200412287D}. In fact, intriguingly, it has been found that some 1D short-range spin models relax the fastest at the QCP \cite{doi:10.1080/00018732.2010.514702,PhysRevLett.103.056403,PhysRevLett.102.130603,2020arXiv200412287D}, contrary to the common intuition of critical slowing down. Dynamical OPs for these spin models also do not exhibit nonanalyticity at the QCP \cite{PhysRevLett.102.130603,Calabrese_2012,2020arXiv200412287D}. 

\begin{figure}
\includegraphics[width=0.48\textwidth]{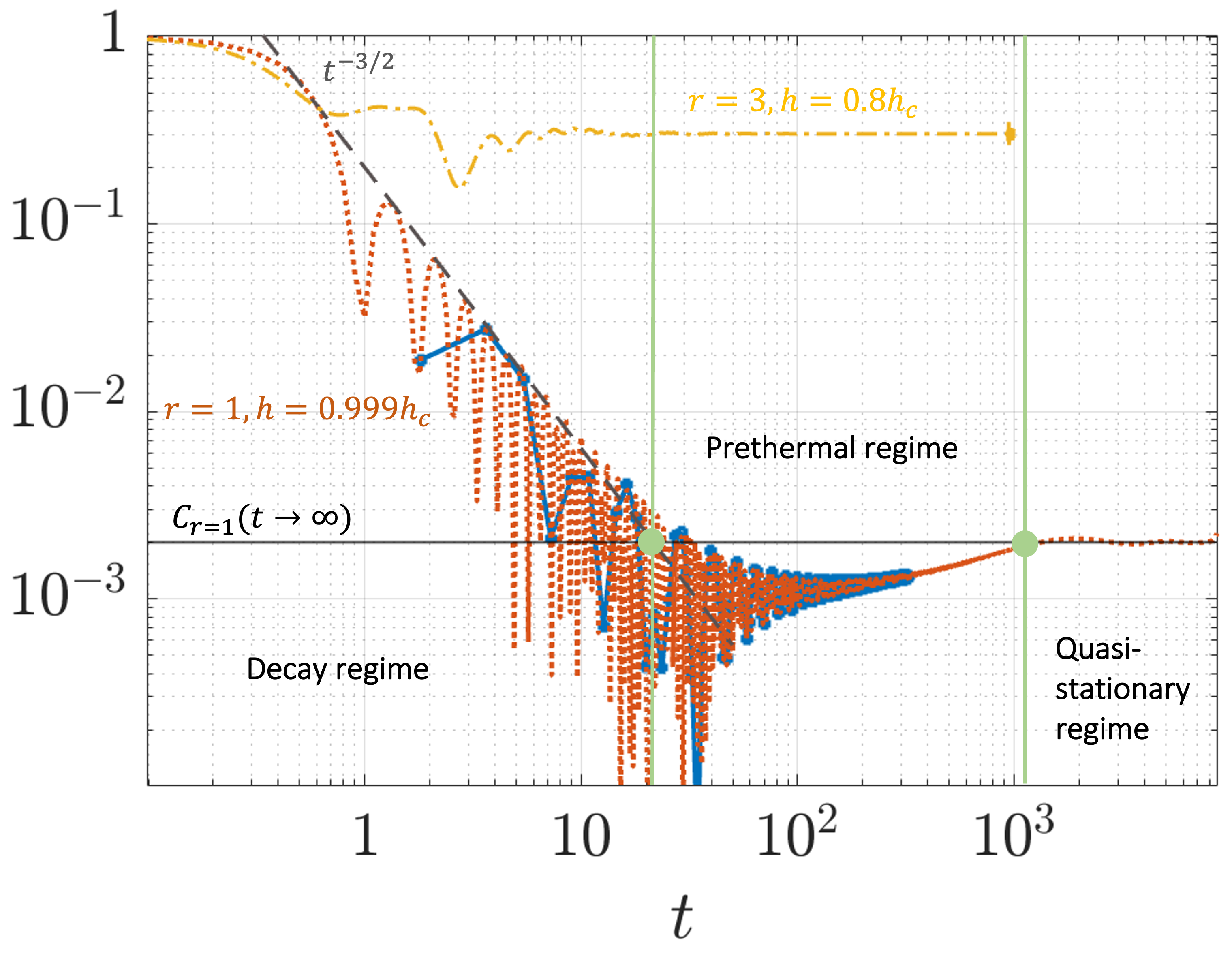}\hfill
\caption{The edge magnetization $C_{1}(t)$ for the Hamiltonian Eq.~\eqref{Hamiltonian} with $\Delta=0$ after a quench in the transverse-field strength from $h_i=0$ to the vicinity of the QCP at $h_c$. The red-dotted curve is plotted based on Eq.~\eqref{series} \cite{1stfootNote}. The blue-squares are values of $|C_{1}(t,h)|$ obtained numerically for the open-boundary TFIC with a system size of $N=1440$, the method of which is detailed in Ref.~\cite{PRBSub}. The panels show the three regimes of time evolution separated by green vertical lines: the decay regime with a power-law decay $\sim t^{-3/2}$ (dashed-gray line) on the left, the prethermal regime in the middle and the quasi-stationary (q.s.) regime on the right. The horizontal black line is $1-h^2$, the q.s.~value. The onsets of prethermal and q.s.~regimes are marked with green balls. As a comparison, the yellow dotted-dashed line plots $|C_{3}(t,h)|$ away from the QCP at $h=0.8h_c$ for $N=1500$ spins and a quench from $h_i=0$ where there is no prethermal regime.}
\label{Fig1}
\end{figure}

In this Letter, we introduce boundaries to short-range 1D spin systems and probe single-site OPs. This reveals a universal prethermal regime upon suddenly quenching to the vicinity of a QCP, when a nonanalyticity of the dynamical OP is present at the QCP. Phrased differently, we show the presence of critical slowing down of OP dynamics near a QCP after a sudden quench. As expected, we find that the duration of the prethermal regime is determined by the inverse energy gap. The universality of the regime holds true for different probe sites, initial conditions, and weak integrability breaking. Further, we analytically and numerically show that the \textit{critically prethermal} regime gives rise to a nonlinear scaling function for the dynamical OP in the reduced control parameter of the QCP. We present our discussion based on a paradigmatic model of QCPs, the transverse-field Ising chain (TFIC).

\textit{Temporal regimes of TFIC.---}~The short-range TFIC with interaction strength $\Delta$ is given by
\begin{eqnarray}
H = -J \sum_{r=1}^{N-1} \sigma_r^z \sigma_{r+1}^z - \Delta  \sum_{r=1}^{N-2} \sigma_r^z \sigma_{r+2}^z + h \sum_{r=1}^N \sigma_r^x, \label{Hamiltonian}
\end{eqnarray}
where $\sigma_r^{x,z}$ are the Pauli spin matrices on sites $r$, $h$ is the transverse-field strength, $N$ is the length of the chain, and we fix $J=1$, which sets the energy scale of the system.  In equilibrium, the TFIC has two phases, i) the ferromagnetically ordered phase for $h<h_c$ and ii) the paramagnetic disordered phase for $h\geq h_c$, where $h_c$ is the QCP. At $\Delta=0$, this model becomes the nearest-neighbor (n.n.) TFIC with a QCP $h_c=1$, and the model is integrable. The QCP shifts to favor order upon introducing interactions with $\Delta > 0$. The OP for this QCP is the magnetization averaged over all sites; when it is finite, it indicates spontaneous symmetry breaking in the ground state and the system is in the ordered phase. 

We consider as initial state the ground state $\Ket{\psi_0}$ of $H$ at initial value $h_i$ of the transverse-field strength, and then we quench the latter to a value $h$. In a periodic chain, the single-site magnetization $C_r(t)=\Bra{\psi_0}\sigma^z_r(t)\Ket{\psi_0}$, at any site $r$, decays exponentially to zero for any $h$ \cite{PhysRevLett.96.136801,Calabrese_2012,PhysRevLett.106.227203,2020arXiv200412287D}, and hence $C_r(t\rightarrow \infty)$ does not host nonanalyticity at the QCP \cite{Calabrese_2012,2020arXiv200412287D}. On the other hand, in an open-boundary chain, $C_r(t)$ stabilizes to a finite nonzero value when $t\rightarrow \infty$ at any $r$ within a finite distance to the boundary, so long as $h_i < h < h_c$. This temporal regime is called quasi-stationary (q.s.) regime \cite{PhysRevLett.106.035701,PRBSub}; see Fig.~\ref{Fig1}. For $h\geq h_c$, $C_r(t \rightarrow \infty)=0$ is suggested by numerical results \cite{PhysRevLett.106.035701,PRBSub} and some analytical arguments \cite{PRBSub}. In our joint submission \cite{PRBSub}, a kink observed at the QCP becomes sharper as the system size increases, and this suggests a nonanalyticity in $C_r(t\rightarrow \infty)$. The origin of this nonanalyticity depends on the presence of zero modes which are induced in the open-boundary chain \cite{PRBSub}. In particular, for the edge magnetization ($r=1$) with $\Delta=0$ and $h_i=0$, there exists a simple analytic form in the infinite time limit $C_{1}(t\rightarrow \infty) = 1-h^2 \equiv C_{1}^{qs}$ for $h<1$ and $C_{1}(t\rightarrow \infty) =0$ for $h\geq 1$ \cite{PhysRevLett.106.035701,Igl_i_2013,PRBSub}. 

\begin{figure*}
\subfloat{\label{fig2a}\includegraphics[width=0.33\textwidth]{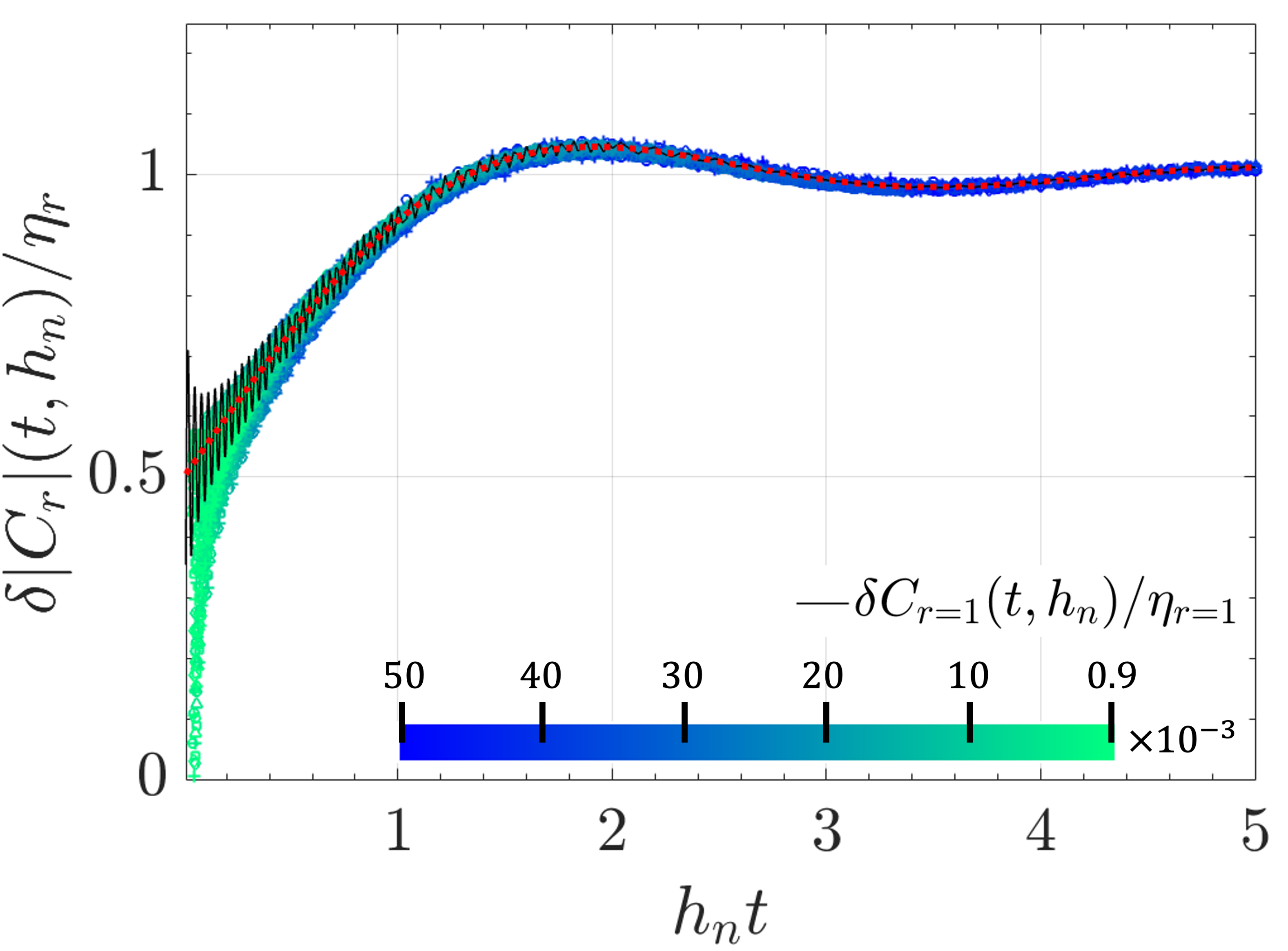}}\hfill
\subfloat{\label{fig2b}\includegraphics[width=0.33\textwidth]{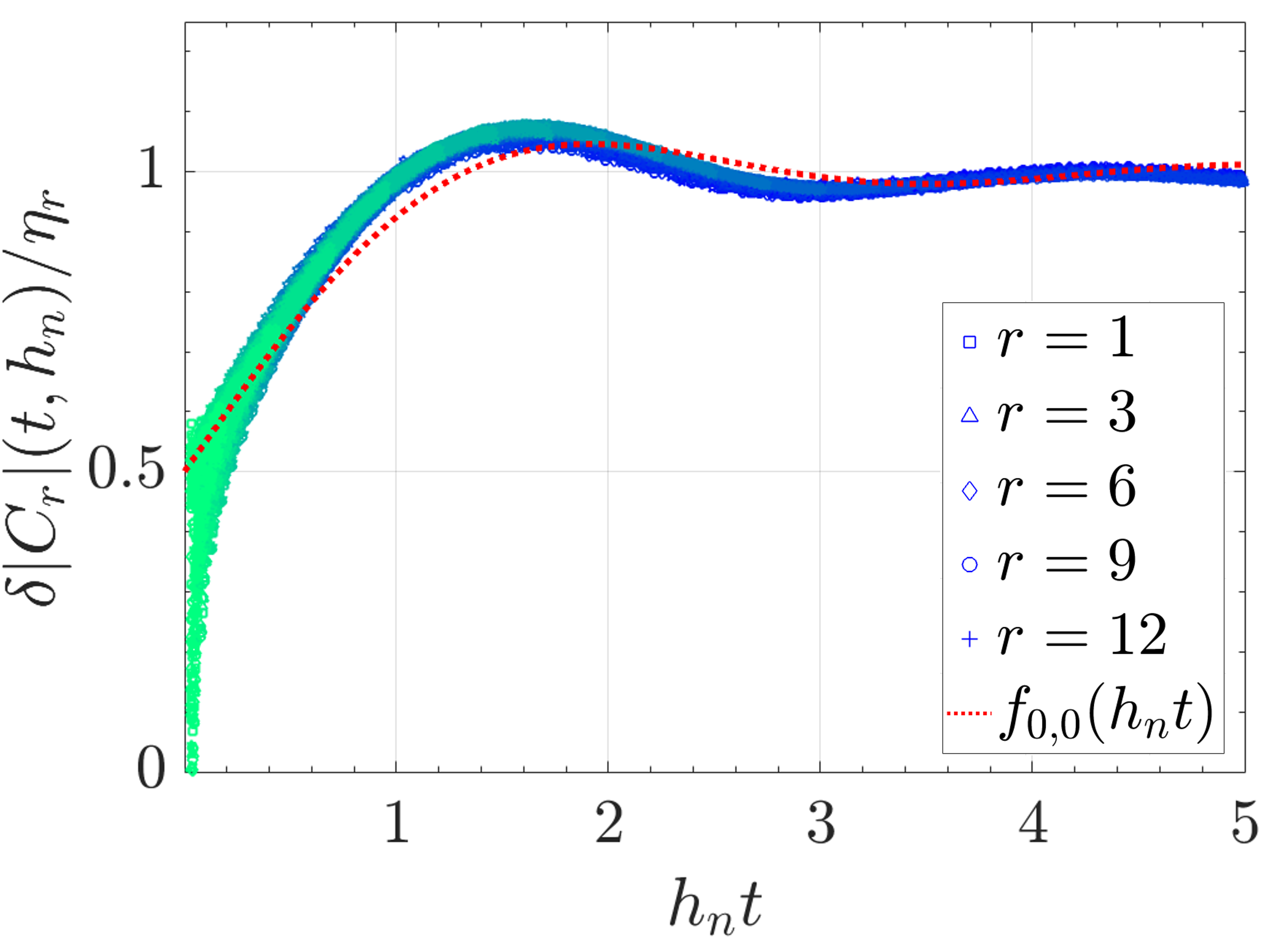}}\hfill
\subfloat{\label{fig2c}\includegraphics[width=0.33\textwidth]{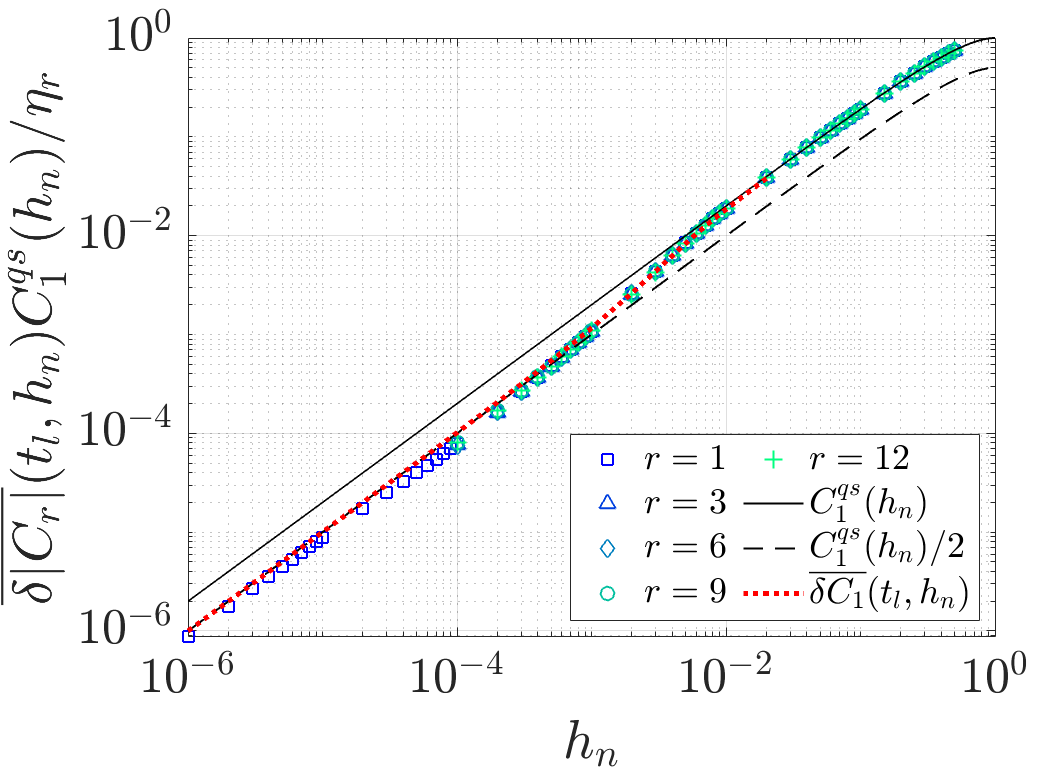}}\hfill
\caption{Scaled critical response $\delta |C_{r}|(t,h_n)/\eta_r$ quenched from $h_i=0$ to $h_n\in [9\times 10^{-4},0.05]$ (colorbar) for (a) the integrable TFIC and (b) near-integrable TFIC, both with a system size of $N=1440$ and for probe sites $r=1,3,6,9,12$. The rescaling factor is $\eta_r=C_{1}^{qs}(h_n) \delta |C_{r}|(t'=280,h)/\delta |C_{1}|(t'=280,h)$ for both (a) and (b). The data in (a) is plotted together with Eq.~\eqref{FullprethermalEq} for $r=1$ (black-solid) and the function $f_{0,0}(h_n t)$ (red-dotted). (c) The dynamical OP for the integrable TFIC with cutoffs $t^*=20$ and $t_l=330$ where data for different $r$ collapse on top of $\overline{\delta  C_{1}}(t_l,h_n)$, Eq.~\eqref{analyticalScaling} (red-dotted). Three different scalings appear corresponding to the different regimes where $t_l$ is in. When $t_l$ is in the decay $(t_l h_n\gg 1)$ or q.s.~$(t_l h_n \ll 1)$~regimes, the data shows linear scalings, described by  $C_{1}^{qs}(h_n)/2$ (dashed-black line) and $C_{1}^{qs}(h_n)$ (solid-black line), respectively. When $t_l$ is in the prethermal regime $(h_n t_l\sim 1)$, the scaling diverges from the two linear functions.} 
\label{Fig2}
\end{figure*}

The single-site magnetization at any $r$ away from the QCP approaches the q.s.~regime as $t^{-3/2}$ after an exponential decay so long as $h_i < h$ \cite{PhysRevLett.106.035701}. Upon quenching to the vicinity of the QCP the decay trend is described only by the power law $t^{-3/2}$. Additionally, an intermediate temporal regime emerges preceding the q.s.~regime (see Fig.~\ref{Fig1})---the nonequilibrium response dips below the q.s.~value and eventually ramps up to it. Figure~\ref{Fig1} shows the time evolution of the edge magnetization $C_{1}(t)$ when the system is quenched from $h_i=0$, e.g.,~$\Ket{\psi_0}=\Ket{\uparrow \uparrow\ldots \uparrow}$ to $h=0.999$, in the integrable (n.n.) TFIC both numerically and analytically \cite{1stfootNote}, where we observe this intermediate regime marked as \textit{prethermal regime}. The onsets $t_{pt}$ and $t_{qs}$ of the prethermal  and q.s.~regimes, respectively, are where the decay roughly ends, i.e.~$t_{pt}^{-3/2} \sim C_{1}^{qs}$, and where a stationary value is attained in the time evolution, respectively (vertical lines in Fig.~\ref{Fig1}). To probe and characterize this prethermal regime, we first define a reduced control parameter $h_n \equiv (h_c-h)/h_c$ as the distance to the QCP and $\delta C_{r}(t,h_n)\equiv C_{r}[t,h=h_c(1-h_n)]-C_{r}(t,h=h_c)$, which we name \textit{critical response}. 
As $h_n \rightarrow 0$, $C_{1}^{qs}(h_n) \approx 2h_n$, we arrive at $t_{pt} \propto h_n^{-2/3}$. The punchline of our paper is that when $h_n\rightarrow 0$ and  $t \gg 1$, the critical response for general $r$  takes the universal form 
\eq{
\delta C_{r}(t,h_n)=C^{qs}_r(|h_n|)f_{\Delta,h_i}(h_n t), \label{eqansatzGeneralrhi}
}
where $f_{\Delta,h_i}(h_n t)$ depends on the weak interaction strength $\Delta$ and the initial condition $h_i$. Note that $C^{qs}_r(|h_n|)$ is the q.s.~value in the ordered phase, while Eq.\ \eqref{eqansatzGeneralrhi} works on both sides of the QCP. Further, $f_{\Delta,h_i}(h_n t)$ is a continuous function of $h_n t $ that satisfies $f_{\Delta,h_i}(h_n t=0)=1/2$ and $f_{\Delta,h_i}(h_n t)=1-f_{\Delta,h_i}(-h_n t)$. When $|h_n| t \gg 1$, $f_{\Delta,h_i}(h_n t)$ approaches $1$ in the ordered phase ($h_n>0$) and approaches $0$ in the disordered phase ($h_n<0$), demonstrating the nonanalyticity in the q.s.~value across the QCP. We plot $f_{\Delta,h_i}(h_n t)$ for $h_i=0$ and $h_n t>0$ in Figs.~\ref{fig2a}~and~\ref{fig2b}, for $\Delta=0$ and $\Delta=0.1$ \cite{2ndfootNote}, respectively.

Eq.\ \eqref{eqansatzGeneralrhi} suggests that  the onset of the the q.s.~regime $t_{qs}\propto h_n^{-1}$, hence the duration of the prethermal regime $\Delta t \equiv t_{qs}-t_{pt}\propto h_n^{-1}$. 
 As the energy of the zero-momentum state in the integrable TFIC is $\epsilon_{k=0} =  h_n$ \cite{sachdev2001quantum}, the prethermal duration $\Delta t \propto \epsilon_{k=0}^{-1}$ is inversely proportional to the single-particle energy gap. The prethermal regime lasts longer as we approach the QCP, motivating the name \textit{critically prethermal regime} and justifying $\delta C_r(t,h_n)$ as the critical response.

In the following, we analytically derive $f_{\Delta,h_i}(h_n t)$ for the edge magnetization at $\Delta=0$ and $h_i=0$, and numerically demonstrate that it holds true for different probe sites $r$. 

\textit{Prethermal regime in the integrable TFIC.---}~The edge magnetization has an analytic series expression whose derivation can be found in \cite{PRBSub},
\begin{eqnarray}
C_{1}(t,h) &=& \sum_{m=0}^{\infty}\frac{(-1)^m}{(2m)!} (2t)^{2m} N_m(h^2), \label{series} \\
N_m(h^2) &=& \sum_{n=1}^m N_{mn}h^{2n}, \hspace{5mm} N_{mn} = \frac{1}{m} \label{eqCr1} \binom{ m}{n-1}\binom{m}{n},\notag
\end{eqnarray}
where $N_m(x)$ are the Narayana polynomials \cite{kostov2009narayana,mansour2009identities}. Eq.~\eqref{series} also describes the two-time edge correlators in the Kitaev chain at infinite temperature \cite{PhysRevB.97.235134}. It has an analytical expression $C_{1}(t,h=1)=J_1(4t)/(2t)$ at the QCP \cite{PRBSub} where $J_1(t)$ is the Bessel function of the first kind.  Additionally, we note that Eq.~\eqref{series} is a generating function of Narayana polynomials and can be expressed  in terms of inverse Laplace transform of a closed form function \cite{supp}. This alternative expression is useful in probing the critically prethermal regime and deriving $f_{\Delta,h_i}(h_n t)$. The critical response in the vicinity of the QCP $h_n \rightarrow 0$ follows \cite{supp}
\begin{subequations}
 \begin{eqnarray}
 \delta C_{1}(t,h_n) &=& C_{1}^{qs}(|h_n|)\left[- \frac{1}{2} J_0(4t) + f(h_n t) \right] \notag \\
 &+&O(h_n^2), \label{FullprethermalEq} \\
 f(h_n t)&\equiv& \frac{1}{2} - \sum_{n=1}^{\infty}\frac{(-h_n t) ^{2n-1}}{(2n)!}\chi_n , \label{ansatzPrethermal}
\end{eqnarray}
\end{subequations}
where $\chi_n\equiv(-1)^{1-n} 2^{1-2 n} (2n-2)!/(n-1)!^2$. $\delta C_1(t,h_n)/C_1^{qs}(h_n)$ for $h_n=0.005$ based on Eq.~\eqref{FullprethermalEq} is plotted as a black-solid line in Fig.~\ref{fig2a}. Here the term $-\frac{1}{2}C_{1}^{qs}(|h_n|) J_0(4t)$, where $J_0(t)$ is the Bessel function of the first kind, introduces oscillations that become negligible when $t \gg 1$. This term also originates from a high frequency expansion in the derivation \cite{supp}, which is why it is only an early-time effect, and hence nonuniversal. The function $f(h_n t)$ can be written in terms of a generalized Hypergeometric function $f(h_n t)= \frac{1}{2}+\frac{(h_nt)}{2}{}_1F_2\left[\left\{\frac{1}{2}\right\};\left\{\frac{3}{2},2\right\};-(h_n t)^2\right]$\cite{supp}, and it is plotted in Fig.~\ref{fig2a} with a dotted-red line. In contrast to the nonuniversal term, $f(h_n t)$ originates from a low frequency ---long-wavelength--- expansion in the derivation, and hence providing extra evidence that the prethermal regime is critical.

Next we demonstrate Eq.\ \eqref{eqansatzGeneralrhi} in the ordered phase using numerics for a finite-size system ($N=1440$). Because our numerics is based on the cluster theorem in the space-time limit \cite{Calabrese_2012}, we obtain numerical values of $|C_r(t,h_n)|$, and hence use $\delta |C_r|(t,h_n)\equiv ||C_r(t,h_n)|-|C_r(t,0)||$ to approximate $\delta C_r(t,h_n)$ \cite{3rdfootNote,PRBSub}. Our numerical data shows that, for $h_n \rightarrow 0$, and $t \gg 1$, $\delta C_{r}(t,h)$ for different choices of $r$ are proportional to each other. Hence defining $\eta_r=C_{1}^{qs}(h_n)\delta |C_r|(t',h_n)/\delta |C_{1}|(t',h_n)$, we found numerically that $\eta_r$ does not depend on $t'$ as long as $t' \gg 1$ \cite{4thfootNote}. For the edge spin, $\eta_{1}=C_{1}^{qs}(h_n)$ by definition. Refs.~\cite{PhysRevLett.106.035701,PhysRevA.69.053616} show that the q.s.~values of the bulk spins have an exponentially decaying spatial profile in  $r$ $\eta_r\approx C_{1}^{qs}(h_n) e^{(r-1)/\xi(h_n)}$, where $\xi(h_n)$ is the correlation length  \cite{supp}.

Fig.~\ref{fig2a} plots $\delta|C_r|(t \geq 50,h_n)/\eta_r$ for all $r=1,3,6,9,12$ quenched from an initial state $h_i=0$ to $h_n\in [ 9\times 10^{-4},0.05]$. The colors, from dark blue to light cyan, correspond to decreasing $h_n$, respectively. The time axis is rescaled by the distance to the QCP, $h_n$. For Fig.~\ref{fig2a}, $t'=280$ is chosen in $\eta_r$. The data collapses on top of each other, and matches well with the analytical function $f(h_n t)$ for $h_n t \gg 0.1$. Therefore, we have numerically demonstrated the validity of Eq.~\eqref{eqansatzGeneralrhi} for different probe sites $r > 1$ in the ordered phase, and hence $f_{0,0}(h_n t)=f(h_n t)$.

\textit{Discussion for $\Delta,h_i \neq 0$.---}~In this section, we discuss Eq.~\eqref{eqansatzGeneralrhi} and $f_{\Delta,h_i}(h_n t)$ for general $\Delta$ and $h_i$. 
We present the case of $\Delta = 0.1$ as an example of the near-integrable model which can be treated with quench mean-field theory (qMFT) \cite{2ndfootNote,PRBSub}. In this case, the QCP is shifted to $h_c \approx 1.165$, and numerical evidence shows that the location of the nonanalyticity observed in the dynamical OP is no longer equal to the QCP \cite{PRBSub}. Hence in~\cite{PRBSub}, we define a dynamical critical point (DCP) based on the nonanalyticity  following Ref.\ \cite{PhysRevLett.123.115701}, and find it to be $h_{dc} = 1.1437$. Since qMFT maps the interacting problem back to a noninteracting problem, we also apply single-particle energy gap analysis in~\cite{PRBSub}, and show that the gap for this noninteracting model indeed closes at $h_{dc} = 1.1437$. Therefore, it is natural to anticipate that a possible critically prethermal regime should emerge around $h_{dc}$ for $\Delta \neq 0$, motivating a  definition of the reduced control parameter as $h_n=(h_{dc}-h)/h_{dc}$. 

Fig.~\ref{fig2b} verifies Eq.\ \eqref{eqansatzGeneralrhi} for $\Delta=0.1$ in the ordered phase using qMFT numerics for  $r=1,3,6,9,12$ quenched from an initial state $h_i=0$ to $h_n\in [8.74\times 10^{-4},0.0557 ]$. Our joint work \cite{PRBSub} shows that for small $\Delta$, $C_{1}^{qs}(h)=\alpha (h_{dc}^{\nu}-h^{\nu})$ where $\alpha$ and $\nu$ are numerically extracted as $\alpha=0.81$ and $\nu=1.81$ for $\Delta=0.1$. Note that for $\alpha=1, \nu=2$ and $h_{dc}=h_c=1$,  we recover the q.s.~value of the edge spin in the integrable TFIC $C_1^{qs}(h)=1-h^2$. Hence, $C_{1}^{qs}(h_n)=\alpha h_{dc}^\nu \left[1-(1-h_n)^\nu\right]$, and we use this expression to define $\eta_1$. $\eta_r$ for $r\neq 1$ are defined similarly  as in the integrable case. Importantly, we find that all data for $\delta|C_r|(t \geq 50,h_n)/\eta_r$ collapse on top of each other, which confirms the validity of Eq.~\eqref{eqansatzGeneralrhi} for small $\Delta \neq 0$. However, the data does not match with the function $f_{0,0}(h_n t)$ (red-dotted line in Fig.~\ref{fig2b}), suggesting that  $f_{\Delta,h_i}(h_n t)$ depends on $\Delta$. In the SM, we verify Eq.~\eqref{eqansatzGeneralrhi} numerically for $h_i\neq 0$ and show that $f_{\Delta,h_i}(h_n t)$ also depends on $h_i$.

For all $\Delta$, $h_i$ and $r$  considered, $C^{qs}_r(|h_n|)\sim |h_n|$ as $h_n\rightarrow 0$.  Specifically, when $h_i=0$, $C^{qs}_{1}(h_n) \approx 2h_n$ for $\Delta=0$ and  $C_{1}^{qs}(h_n) \approx \alpha\nu h_{dc}^\nu h_n $  for $\Delta=0.1$. The case of $h_i \neq 0$ is discussed in Ref.\ \cite{PRBSub}. The linear scaling of $C^{qs}_r(|h_n|)$ in $h_n$ results in the self-similarity of the critical response: When $h_n\rightarrow 0$, $t \gg 1$ and $ \kappa t \gg 1$, $\delta C_{r}(t,h_n)=\kappa \delta C_{r}(\kappa^{-1}t,\kappa h_n)$ where $\kappa$ is a rescaling factor.

\textit{Scaling of dynamical OP near QCP.---}~
Finally, we probe the critically slowed down prethermal regime in the ordered phase $(h_n>0)$ by studying the scaling of a dynamical OP defined with a finite long-time cutoff $t_l$:
\begin{eqnarray}
\overline{\delta C_r}(t_l,h_n) &\equiv & \frac{1}{t_l-t^*}\int_{t^*}^{t_l} dt\ \delta C_r(t,h_n),\label{DOP}
\end{eqnarray}
where $t^*$ is a short-time cutoff with negligible influence on the value of $\overline{\delta C_r}(t_l,h_n)$ \cite{supp}.
This newly introduced dynamical OP extends beyond the current paradigm of probing the dynamical scaling near a QCP at infinite time, and   enables the discussion of experiments often limited by finite coherence times. Here we can imagine $t_l$ as an experimentally (or computationally) the longest time accessible. The temporal cutoff can be extended to $t_l \rightarrow \infty$ if desired. 

When $t^*=0$,   Eq.\ \eqref{ansatzPrethermal} together with Eq.\ \eqref{eqansatzGeneralrhi}  suggest that the dynamical OP  for $\Delta=0$ and $h_i=0$ is given by \cite{5thfootNote}
\begin{eqnarray}
\overline{\delta C_{r}}(t_l,h_n) &=& C_r^{qs}(|h_n|)\left[ \frac{1}{2}-\sum_{n=1}^{\infty}\frac{(-h_n t_l)^{2n-1}}{2n\times (2n)!}\chi_n\right] \notag \\
&+&O(h_n t_l^{-1})+O(h_n^2). \label{analyticalScaling}
\end{eqnarray}
$\overline{\delta C_{r}}(t_l,h_n) $ for $r=1$ is plotted in Fig.~\ref{fig2c} as the red-dotted line for $t_l=330$. When $t_l\gg 1$ and $h_n\rightarrow 0$, the first line of Eq.\ \eqref{analyticalScaling} gives a good approximation of $\overline{\delta C_{r}}(t_l,h_n)$.  For  $h_n t_l \ll 1$ and $h_n t_l \gg 1$, $t_l$ probes the beginning of the prethermal ramp and the q.s.~regime, respectively. In these regimes, we observe $\overline{\delta C_{1}}(t_l,h_n) \approx \frac{1}{2}C^{qs}_1(h_n)$ (dashed-black) and $\overline{\delta C_{1}}(t_l,h_n) \approx C^{qs}_1(h_n)=1-(1-h_n)^2$ (solid-black), respectively. Both are linear in $h_n$ for $h_n\ll 1$, and connected through a nonlinear crossover when $h_n t_l\sim 1$ holds, and hence when $t_l$ probes the prethermal ramp.

Similar to the previous discussion, we numerically define $\overline{\delta |C_r|}(t_l,h_n)$ as the time average of $\delta |C_r|(t,h_n)$ between  $t^*$ and $t_l$. To demonstrate that the dynamical OP has a similar scaling behavior for different $r$, we rescale the data using $\eta_r$ and plot $\overline{\delta |C_r|}(t_l,h_n)C^{qs}_1(h_n)/\eta_r$ in Fig.~\ref{fig2c}. Note that $\overline{\delta |C_r|}(t_l,h_n)C^{qs}_1(h_n)/\eta_r = \overline{\delta |C_r|}(t_l,h_n)$ for $r=1$ by definition. The linear-to-linear crossover in $\overline{\delta C_r}(t_l,h_n)$ for small $h_n>0$, demonstrated  in Fig.~\ref{fig2c}, is universal for any $\Delta$ and $h_i$, and robust against changing $t_l$ \cite{supp}, while the shape of the nonlinear crossover depends on $f_{\Delta, h_i}(h_nt)$. This is suggested by Eq.\ \eqref{eqansatzGeneralrhi}, where $f_{\Delta, h_i}(h_n t)$ has universal limiting properties  and  $C^{qs}_r(h_n)$ always has linear scaling in $h_n$. To demonstrate the universality, we plot the numerical data of $\overline{\delta |C_r|}(t_l,h_n)C^{qs}_1(h_n)/\eta_r$ for $\Delta =0.1, h_i=0$ and $\Delta=0, h_i = 0.1$ in the SM. 

\textit{Conclusion and outlook.---}~
We  discover critical slowing down in the open-boundary TFIC upon suddenly quenching to the vicinity of the QCP. This critical slowing down is expressed in Eq.~\eqref{eqansatzGeneralrhi} universally for any probe site, weak interactions or the initial state, and rigorously proven for a special case. Analytical analysis leads us to reveal self-similarity in the dynamics and find that the duration of the prethermal regime diverges as one approaches to the QCP because of the gap closing. An immediate next step is to confirm the presence of duality across the QCP, in other words the applicability of Eq.~\eqref{eqansatzGeneralrhi} in the disordered phase. An interesting question to answer is whether Eq.~\eqref{eqansatzGeneralrhi} is applicable to nonintegrable TFIC beyond qMFT. Finally, it is worth checking whether other order parameters \cite{PhysRevLett.123.115701} and other systems that host dynamical phase transitions after sudden quenches \cite{Mori_2018} could also exhibit the critically prethermal regime.

We thank S.~Yelin for helpful discussions. C.B.D.~is supported by the ITAMP grant at Harvard University. Y.W.~is supported by AFOSR. J.C.H.~and P.U.~acknowledge support by Provincia Autonoma di Trento and the ERC Starting Grant StrEnQTh (project ID 804305), Q@TN — Quantum Science and Technology in Trento — and the Collaborative Research Centre ISOQUANT (project ID 273811115). 
\bibliographystyle{apsrev4-1}
%\bibliography{Bibliography} 

\begin{thebibliography}{50}
\bibitem [{1st()}]{1stfootNote}%
  \BibitemOpen
  \href@noop {} {\bibinfo  {journal} {$\delta C_{1}(t,h_n)/C_1^{qs}(h_n)$ in
  Fig.~\ref{Fig1} can be plotted using either an alternative expression for
  Eq.\ \eqref{series} or the inverse Laplace transform expression detailed in
  \cite{supp}, as evaluating the series summation in Eq.\ \eqref{series} for
  long times becomes virtually impossible}\ }\BibitemShut {NoStop}%
\bibitem [{2nd()}]{2ndfootNote}%
  \BibitemOpen
  \href@noop {} {\bibinfo  {journal} {The near-integrable model is mapped to
  the single-particle picture and the dynamics are calculated via quench
  mean-field theory (qMFT). qMFT introduces next n.n.~coupling and pairing to
  the usual Kitaev chain and simply renormalizes the n.n.~coupling and pairing
  strengths $J$ as well as the chemical potential $h$. Hence, the QCP shifts
  favoring order. For details of this method and the results, see
  \cite{PRBSub}}\ }\BibitemShut {NoStop}%
\bibitem [{3rd()}]{3rdfootNote}%
  \BibitemOpen
  \href@noop {} {\bibinfo  {journal} {This is a good approximation when
  $C_r(t,h_n)$ and $C_r(t,0)$ are both positive, hence it is expected to work
  well only in the ordered phase. This approximation is also the reason why we
  observe a mismatch between numerics and analytics in Fig.~\ref{fig2a} for the
  values $h_n t \ll 1$}\ }\BibitemShut {NoStop}%
\bibitem [{4th()}]{4thfootNote}%
  \BibitemOpen
\bibfield  {journal} {  }\href@noop {} {\bibinfo  {journal} {This indicates
  that when $t' \geq t_{qs}$, $\eta_r=\delta C_r(t=\infty,h_n)
  =C^{qs}_{r}(h_n)$ can be used to numerically extract the q.s. value of spin
  $r$ from finite-time data [using the fact that $C_{r}(t=\infty, h=1)=0$]}\
  }\BibitemShut {NoStop}%
\bibitem [{5th()}]{5thfootNote}%
  \BibitemOpen
  \href@noop {} {\bibinfo  {journal} {Eq.~\eqref{analyticalScaling} can also be
  expressed as a Hypergeometric function \cite{supp}}\ }\BibitemShut {NoStop}%
\end{thebibliography}


%\newpage

\onecolumngrid
%\newpage

\appendix
\setcounter{equation}{0}
\setcounter{figure}{0}
\renewcommand{\thetable}{S\arabic{table}}
\renewcommand{\theequation}{S\thesection.\arabic{equation}}
\renewcommand{\thefigure}{S\arabic{figure}}
\setcounter{secnumdepth}{2}

\begin{center}
{\Large Supplementary Material \\ 
\vspace{0.2cm}
%\titleinfo
Probing dynamical criticality near quantum phase transitions
}
\end{center}
\section{Deriving the critical response function for $\Delta=h_i=0$}
In this section, we give an alternative expressions of Eq.~\eqref{series}  and derive the critical response function for the edge magnetization when $h_i=0, \Delta=0$. The main idea is to calculate $C_1(t,h)$ as a special class of generating functions of Narayana polynomials $N_m(h^2)$ \cite{kostov2009narayana,mansour2009identities}. 

First of all, we define two generating functions for $N_m(h^2)$:
\eq{
\theta(\tau,h^2)&\equiv  \sum_{m=0}^{\infty}\frac{1}{(2m)!} \tau^m N_m(h^2), \label{seriesfSupp1} \\
g(\tau,h^2)&\equiv  \sum_{m=0}^{\infty} \tau^m N_m(h^2)\label{seriesGSupp1},\\
&=\frac{1}{2\tau}\left[1+\tau(1-h^2)-\sqrt{1-2\tau(h^2+1)+\tau^2(h^2-1)^2}\right]\label{seriesGSupp2}.
}
Here, $g(\tau,h^2)$ is called the ordinary generating functions of $N_m(h^2)$ and it has a closed-form expression given by Eq.\ \eqref{seriesGSupp2}.
From Eq.~\eqref{series} in the main text, we see that $C_1(t,h)=\theta(-4t^2,h^2)$. The ordinary generating function $g(\tau,h^2)$ and the generating function in Eq.\ \eqref{seriesfSupp1} has the following relation
\eq{
g(-\tau,h^2)=\int_0^{\infty} e^{-s}\theta(-s^2\tau,h^2)ds. \label{relation}
}
Hence we can calculate $\theta(-\tau,h^2)$ using the expression of $g(-\tau,h^2)$ [Eq.\ \eqref{seriesGSupp2}] and the relation between $\theta(\tau,h^2)$ and $g(\tau,h^2)$ [Eq.~\eqref{relation}].

Defining $\Theta(s,h^2)=\theta(-s^2,h^2)$ for $s>0$, we have,  %\int_0^{\infty} e^{-s}F(s\sqrt{\tau},h^2)ds=
\begin{eqnarray}
g(-\tau,h^2)=\frac{1}{\sqrt{\tau}}\int_0^{\infty} e^{-s/\sqrt{\tau}}\Theta(s,h^2)ds.
\end{eqnarray}
Hence $\sqrt{\tau} g(-\tau,h^2)$ is the Laplace transform of $\Theta(s,h^2)$ from $s$ to $1/\sqrt{\tau}$: $\sqrt{\tau} g(-\tau,h^2) =\mathcal{L}[\Theta(s,h^2),s, \frac{1}{\sqrt{\tau}}]$. Therefore, defining $y=1-h^2$ for this section, $C_1(t,h^2)=\Theta(2t,h^2)$ can be written as the inverse Laplace transform of the function $G(\tau)\equiv \tau^{-1} g(-\tau^{-2},h^2)$:
\begin{subequations}
\eq{
C_1(t,h^2)&=\Theta(2t,h^2)=\mathcal{L}^{-1}[G(\tau,y),\tau,2t]\\
G(\tau)&=\frac{y}{2\tau}-\frac{\tau}{2}+\frac{1}{2}\sqrt{\tau^2+4-2y+\left(\frac{y}{\tau}\right)^2},
}
where $\tau$ can be regarded as frequency. 
\label{eqSLaplaceT}
\end{subequations}
Eq.\ \eqref{eqSLaplaceT} 
 At the critical point, $h=1$, $y=0$,
\eq{
C_1(t,h^2)=\mathcal{L}^{-1}\left[-\frac{\tau}{2}+\frac{1}{2}\sqrt{\tau^2+4},\tau,2t\right]= \frac{J_1(4t)}{2t},
}
which has been given in the main text. The critical response parameterized by the control order parameter $h_n=1-h$ is given by 
\eq{
\delta C_1(t,h_n)\equiv C_1(t,h^2)-C_1(t,1)=\mathcal{L}^{-1}\left[
\Delta G(\tau,y),\tau,2t\right] \label{eqsuppcriticalresponseinv}
}
where
\eq{
\Delta G(\tau,y)\equiv G(\tau,y)-G(\tau,0)=\frac{y}{2\tau}+\frac{1}{2}\sqrt{\tau^2+4-2y+\left(\frac{y}{\tau}\right)^2} -\frac{1}{2}\sqrt{\tau^2+4}. \label{eqsuppDeltaG}
}
Since we want to study the critical response close to the QCP, we focus on the case $|y|\ll 1$. 
In the following, we introduce an approximate expression of $\Delta G(\tau,y)$ applicable at all $\tau$, motivated by both the low and high frequency expansion of $\Delta G(\tau,y)$. 
%For this reason, we will expand Eq.~\eqref{eqsuppDeltaG} both in high and low frequency. 
(i) High frequency expansion: when $\tau\gg |y|^{1/2}$,  $(y/\tau)^{2}\ll|2y| \ll \tau^2+ 4$ under the square root of the second term in Eq.\ \eqref{eqsuppDeltaG},
\eq{
\Delta G(\tau, y)=\frac{y}{2\tau}-\frac{1}{4}\frac{2y}{\sqrt{\tau^2+4}}+o(y). \label{eqsuppDeltaG1}
}
Here, $o(y)$ is the little-o notation. $f(y)=o(y)$ implies $\lim_{y\rightarrow 0} f(y)/y=0$. 

(ii) Low-frequency expansion: when $\tau\ll |y|^{1/2}$,  $|\tau^2-2y|\ll (y/\tau)^2$,  $4+(y/\tau) ^{2}$ is the dominant term in the square root of the second term in Eq.\ \eqref{eqsuppDeltaG}, 
\eq{
\Delta G(\tau, y)=\frac{y}{2\tau}+\frac{1}{2}\sqrt{4+\left(\frac{y}{\tau}\right)^2}-1+ o(y).\label{DifferenceG}
}

Motivated by Eqs.\ \eqref{eqsuppDeltaG1} and  Eqs.\ \eqref{DifferenceG}, we numerically verify the following expression for all values of $\tau>0$,
\eq{
\Delta G(\tau, y)=-\frac{y}{2\sqrt{\tau^2+4}} + \left[\frac{y}{2\tau}+\sqrt{1+\frac{1}{4}\left(\frac{y}{\tau}\right)^2}-1\right] +O(y^{2}). \label{eqsuppDeltaG3}
}
Follow Eq.\ \eqref{eqsuppcriticalresponseinv} by applying the inverse Laplace transform to Eq.\ \eqref{eqsuppDeltaG3}, we obtain
\eq{
\delta C_1(t,h_n)=-\frac{y}{2} J_0(4t)+y\left[\frac{1}{2}+\frac{yt}{4}{}_1F_2\left(\left\{\frac{1}{2}\right\};\left\{\frac{3}{2},2\right\};-\frac{(yt)^2}{4}\right)\right]
+O(y^{2}),\label{eqsuppcriticalresponsey}
}
where the first two terms  describe the nonuniversal short-time oscillation and the universal critical slowing down, respectively. We see that the small $\tau$ behavior of $\Delta G(\tau, y)$ decides the self-similar dynamics in the prethermal and q.s.~regimes. Note that the Laplace transform of the first term in Eq.~\eqref{eqsuppcriticalresponsey} corresponds to the first term in Eq.~\eqref{eqsuppDeltaG3}, and consistently this term originates from the high-frequency expansion. Whereas the the rest of the terms in Eq.~\eqref{eqsuppcriticalresponsey} originates from the low frequency expansion. Note that these terms are also the terms that describe the universal critical dynamics of prethermal and q.s.~regimes. Hence, it is consistent with the fact that the critical dynamics originate from low-frequency (long-wavelength) physics.

Since $y=2h_n+O(h_n^2)$ as $h\rightarrow 1$, we can rewrite the r.h.s of Eq.\ \eqref{eqsuppcriticalresponsey} as a function of $h_n$:
\eq{
\delta C_1(t,h_n)=2h_n 
\left[ -\frac{1}{2} J_0(4t)+ \frac{1}{2}+\frac{h_nt}{2}{}_1F_2\left(\left\{\frac{1}{2}\right\};\left\{\frac{3}{2},2\right\};-(h_n t)^2\right)\right]
+O(h_n^{2}).\label{criticalResponseClosedForm}
}
This equation is identical to the Eq.~\eqref{FullprethermalEq} in the main text for $\Delta=0$ and $h_i=0$, where the series expansion in $f(h_n t)$ corresponds to the generalized Hypergeometric function term in Eq.\ \eqref{criticalResponseClosedForm}.

The dynamical order parameter for $\Delta=0,h_i=0$,  $h_n>0$ and $t^*=0$ [Eq.\ \eqref{analyticalScaling} in the main text] can be written as 
\begin{eqnarray}
\overline{\delta C_{r}}(t_l, h_n) &=& C^{qs}_r(h_n)\left[
\frac{1}{2} +\frac{1}{4h_n t_l} J_0(2h_nt_l)+\frac{h_n t_l}{2}\ {}_1F_2\left(\left\{\frac{1}{2}\right\};\left\{\frac{3}{2},2\right\};-(h_n t_l)^2\right)\right]+O(h_n t_l^{-1})+O(h_n^2) \notag
\end{eqnarray}
where $O(h_n t_l^{-1})$ comes from the non-universal initial oscillations in $\delta C_{r}(t, h_n) $.

\section{Recasting Eq.~\eqref{series} as a Taylor expansion around the QCP}

While each term in the series expansion in Eq.~\eqref{series} diverges when $t$ increases, we expect that the cancellation between these large numbers results in a number with absolute value between $0$ and $1$.  Since any numerical simulation has finite digits of accuracy for numbers, evaluating the series summation properly at large $t$ becomes virtually impossible. This numerical challenge poses a serious problem to the study of quench dynamics close to the critical point, where we need to probe an increasing duration of time due to the critical slowing down.

In the following, we propose yet another method of recasting Eq.~\eqref{series} using the observation that derivatives of $C_{1}(t,h)$ with respect to $h^2$ at $h=1$ have closed-form expressions. We define $k=h^2$ for convenience for the rest of this section.  $C_1(t,h)$ is differentiable with respect to $k$ at $k=1$ and its $n$-th order derivative is given by 
\eq{
 \partial^n_{k} C(t,k=1)&=\sum_{m=n}\frac{(-1)^{m}(2m-n)!}{(2m)!(m-n+1)!(m-n)!}(2t)^{2m}\notag \\
 &=(-1)^n (2t)^{2n}\frac{n!}{(2n)!}{}_2F_3[\{\frac{n+1}{2},\frac{n+2}{2}\},\{2,\frac{1}{2}+n, n+1\}, -4t^2].\notag
 }
Therefore, through Taylor expansion around the QCP, the edge magnetization reads,
\begin{eqnarray}
C_{1}(t,h)&=&\sum_{n=0}^{\infty}\frac{(h^2-1)^n}{n!}\partial^n_k C(t,k=1), \\
&=& \sum_{n=0}^{\infty}\frac{(1-h^2)^n}{(2n)!} (2t)^{2n}{}_2F_3[\{\frac{n+1}{2},\frac{n+2}{2}\},\{2,\frac{1}{2}+n, n+1\}, -4t^2]. \label{seriesExpansion}
\end{eqnarray}
Eq.~\eqref{seriesExpansion} is utilized to plot the analytical time evolution in Fig.~1 and the solid-black line in Fig.~2a in the main text with around $40$ terms. Eq.~\eqref{seriesExpansion} can be proven with the help of the following lemmas. 
\begin{lemma} \label{theoremCderiv}
\eq{
\partial_k^{n}C(t,k)=\sum_{m=0}^{\infty}\frac{(-1)^m}{(2m)!} (2t)^{2m} \partial_k^{n}N_m(k). \label{eqpartialynC}
}
\end{lemma}
\begin{lemma}\label{LemmaderiN}
For integer $n\geq 0$, 
\eq{
\partial^n_k N_m(k)|_{k=1}
&
=\begin{cases}
\frac{(2m-n)!}{(m-n+1)!(m-n)!}& m\geq n\\
0 &m=0,1,\dots n-1
\end{cases}.\label{eqpartynN}
%=\sum_{l=n}^m\prod_{j=0}^{n}(l-j)\frac{1}{m} \binom{ m}{l-1}\binom{m}{l}
}
\end{lemma}
Hence the proof of Eq.~\eqref{seriesExpansion} is reduced to the proofs of Lemma \ref{theoremCderiv} and \ref{LemmaderiN}. We first do the proof of Lemma \ref{LemmaderiN}.  After that,  Lemma \ref{theoremCderiv} will be proved using  Lemma \ref{LemmaderiN}.
\begin{proof}(Lemma \ref{LemmaderiN})
Taking the derivatives of the Narayana polynomials  $N_m(k)$, we have
\eq{
 \partial^n_{k}N_m(k)&= \sum_{l=n}^m \prod_{j=0}^{n-1} (l-j)N_{m,l}k^{l-n}= \sum_{l=n}^m \frac{l!}{(l-n)!}N_{m,l}k^{l-n},
 }
Setting $k=1$, we have
\eq{
\partial^n_{k}N_m(k=1)&=\sum_{l=n}^m \frac{l!}{(l-n)!} \frac{1}{m} \binom{ m}{l-1}\binom{m}{l},\\
&=\begin{cases}
\frac{(2m-n)!}{(m-n+1)!(m-n)!}& m\geq n\\
0 &m=0,1,\dots n-1
\end{cases},
}
which concludes the proof for Lemma \ref{LemmaderiN}.
\end{proof}
We continue to prove Lemma \ref{theoremCderiv}. 
\begin{proof}(Lemma \ref{theoremCderiv})
For convenience, we define $\omega_m(k)=\frac{(-1)^m}{(2m)!} (2t)^{2m} N_m(k)$. Hence $C(t,k)=\sum_{m=0}^{\infty} \omega_m(k)$.  
According to  Tannery's theorem in mathematical analysis, if  $\sum_{m=0}^{\infty} \partial_k^n \omega_m(k)$ is uniformly convergent,  we have
\eq{
\sum_{m=0}^{\infty}\partial_k^{n}\omega_m(k)=\partial_k^{n}\sum_{m=0}^{\infty}\omega_m(k)=\partial_k^{n}C(t,k), \label{equniformconverge}
}
which is the objective of our proof. 
In the following, we prove the uniform convergence of $\sum_{m=0}^{\infty}\partial_k^{n}\omega_m(k)$.

For any given $n$ and $t$, we have
\eq{
\lim_{m\rightarrow \infty}\left|\frac{\partial_k^n \omega_{m+1}(t)}{\partial_k^n \omega_m(t)}\right|=\lim_{m\rightarrow 0}\frac{(2m-n+1)(2m-n+2)}{(m+1)(2m+1)(2m+2)(m-n+1)(m-n+2)}t^2 =0, \label{eqratiolimit}
}
where in the first equality we have used Lemma \ref{LemmaderiN}.
According to the Weierstrass M-test for uniform convergence and the ratio test for series convergence, Eq.\ \eqref{eqratiolimit} implies that the series $\sum_{m=0}^{\infty}\partial_k^{n}\omega_m(k)$  is uniformly convergent, and hence Eq.\ \eqref{equniformconverge} holds. This is the end of the proof for Lemma \ref{theoremCderiv}.

\end{proof}

\section{Critical response for $h_i\neq 0$}

\begin{figure}
\subfloat{\label{supFig1a}\includegraphics[width=0.5\textwidth]{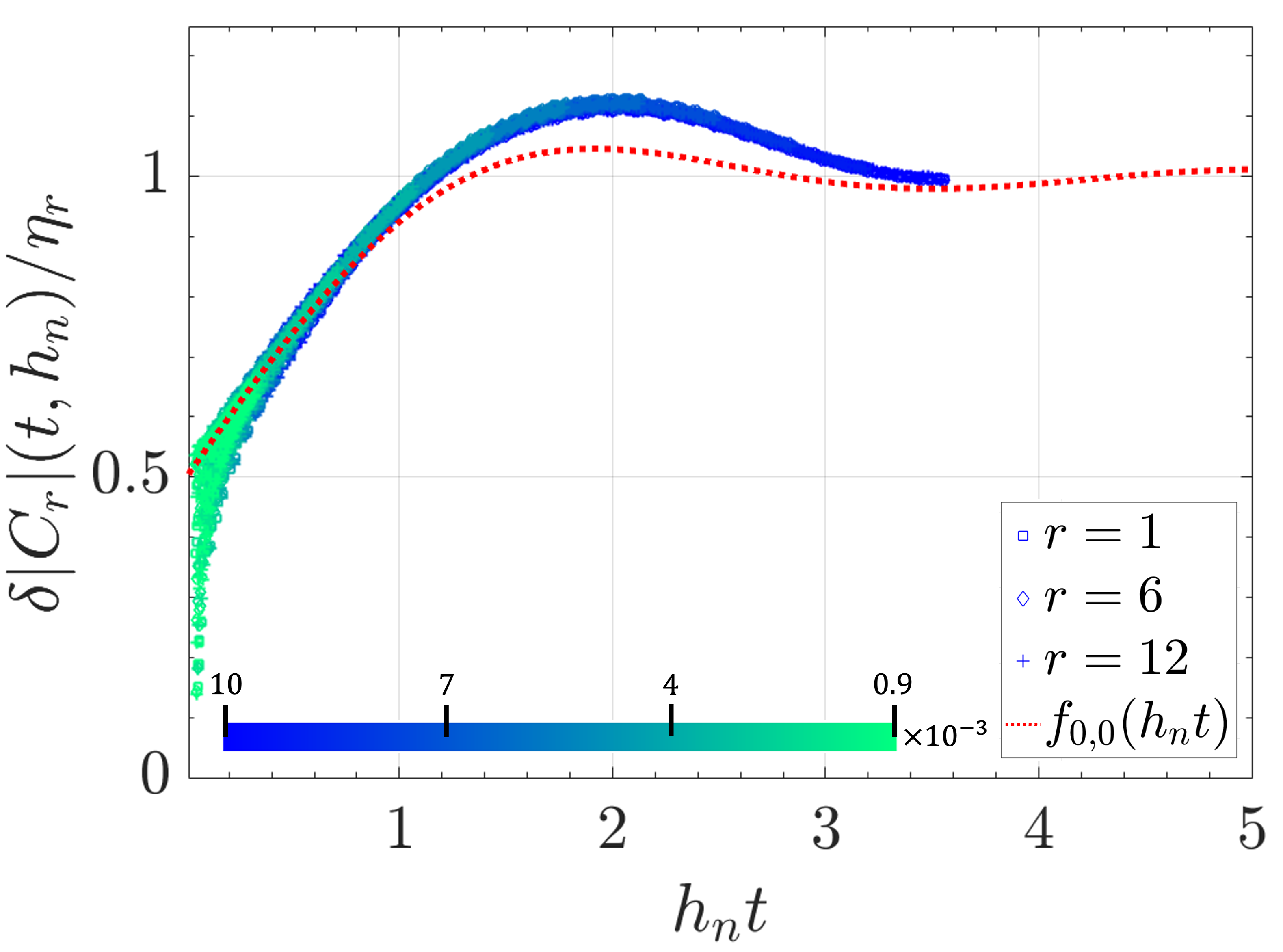}}\hfill
\subfloat{\label{supFig1b}\includegraphics[width=0.5\textwidth]{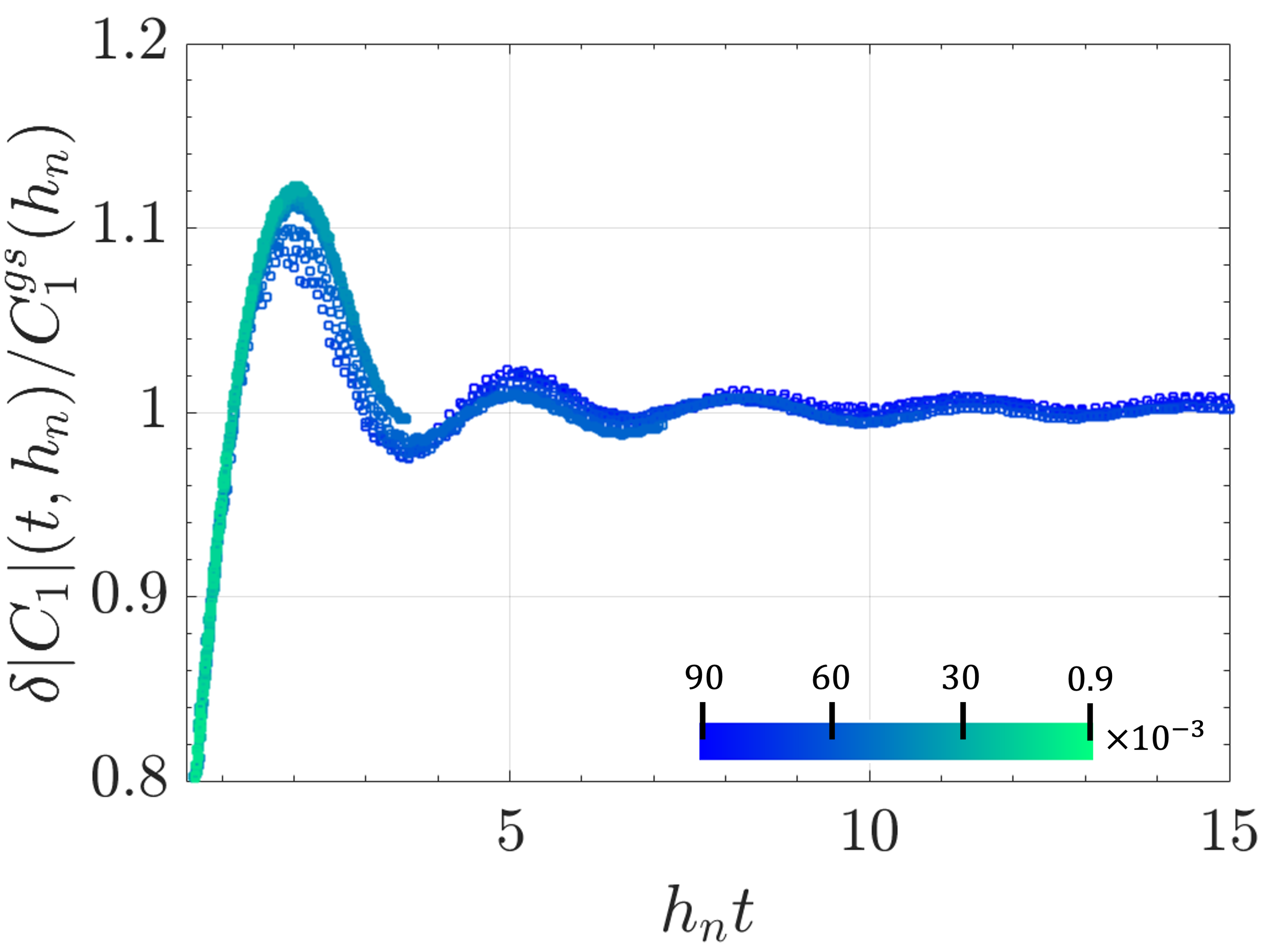}}\hfill
\caption{Scaled critical response $\delta |C_{r}|(t,h_n)/\eta_r$ quenched from $h_i=0.1$ to $h_n\in [9\times 10^{-4},0.01]$ (colorbar) for the integrable TFIC with a system size of $N=1440$ and for probe sites $r=1,6,12$. The rescaling factor is $\eta_r=C_{1}^{qs} \delta |C_{r}|(t=300,h)/\delta |C_{1}|(t=300,h)$. The data in (a) is plotted against the function $f_{0,0}(h_n t)$. (b) The same system with (a) but only for the edge spin $r=1$ and quenched to $h_n\in [9\times 10^{-4},0.09]$ (colorbar).  The plot shows data far away from the QCP not matching perfectly with the data close to QCP.}
\end{figure}

In this section, we demonstrate that our main result, i.e. Eq.~\eqref{eqansatzGeneralrhi} in the main text, works for general initial conditions using an example of $h_i=0.1$. Fig.~\ref{supFig1a} plots $\delta|C_r|(t,h_n)/\eta_r$ quenched to $h_n \in [9\times 10^{-4},0.01]$ for probe sites $r=1,6,12$ and compared to the function $f_{0,0}(h_n t)$. All data points to a single function for the critical response $f_{0,0.1}(h_n t)$ and this function is different from $f_{0,0}(h_n t)$. We also plot the rescaled critical response when the system is quenched to a larger interval of $h_n \in [9\times 10^{-4},0.09]$ in Fig.~\ref{supFig1b}. We observe that the critical response is able to differentiate the distance to the critical point. Data for $h_n > 0.01$ does not match perfectly, which suggests that the region $h_n > 0.01$ is seen as away from the QCP when $h_i=0.1$ is set.

\begin{figure*}
\subfloat{\label{supFig5a}\includegraphics[width=0.33\textwidth]{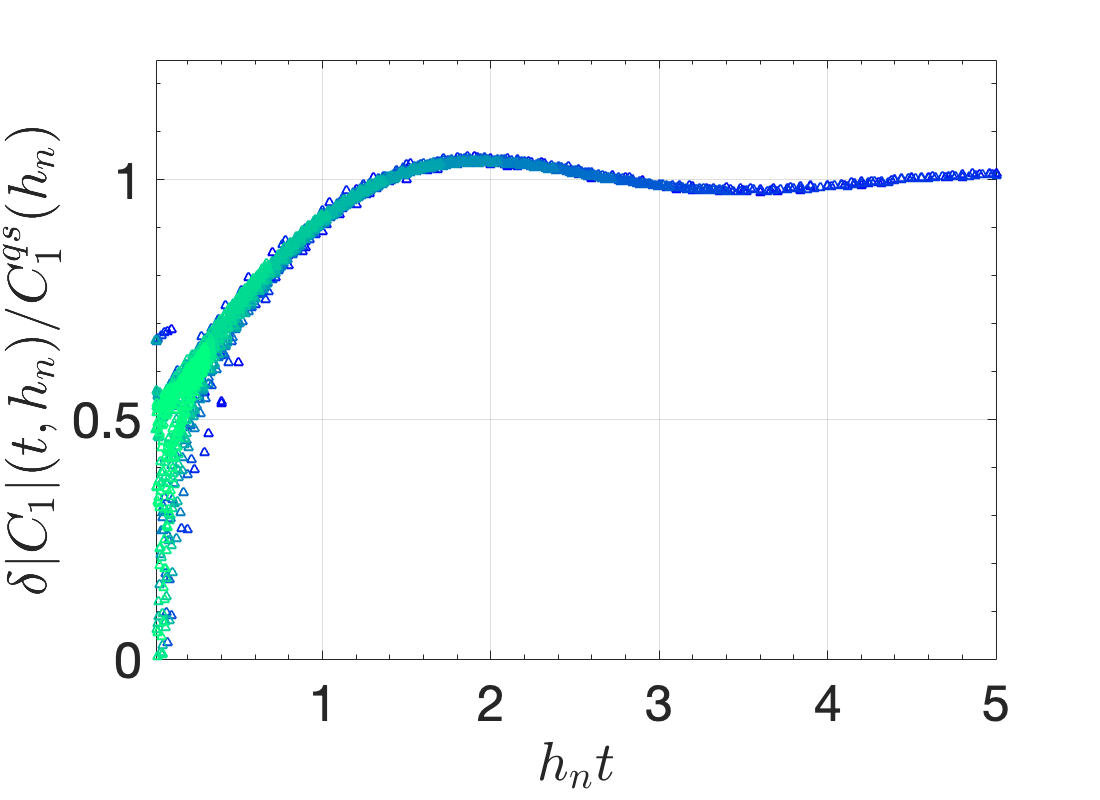}}\hfill
\subfloat{\label{supFig5b}\includegraphics[width=0.33\textwidth]{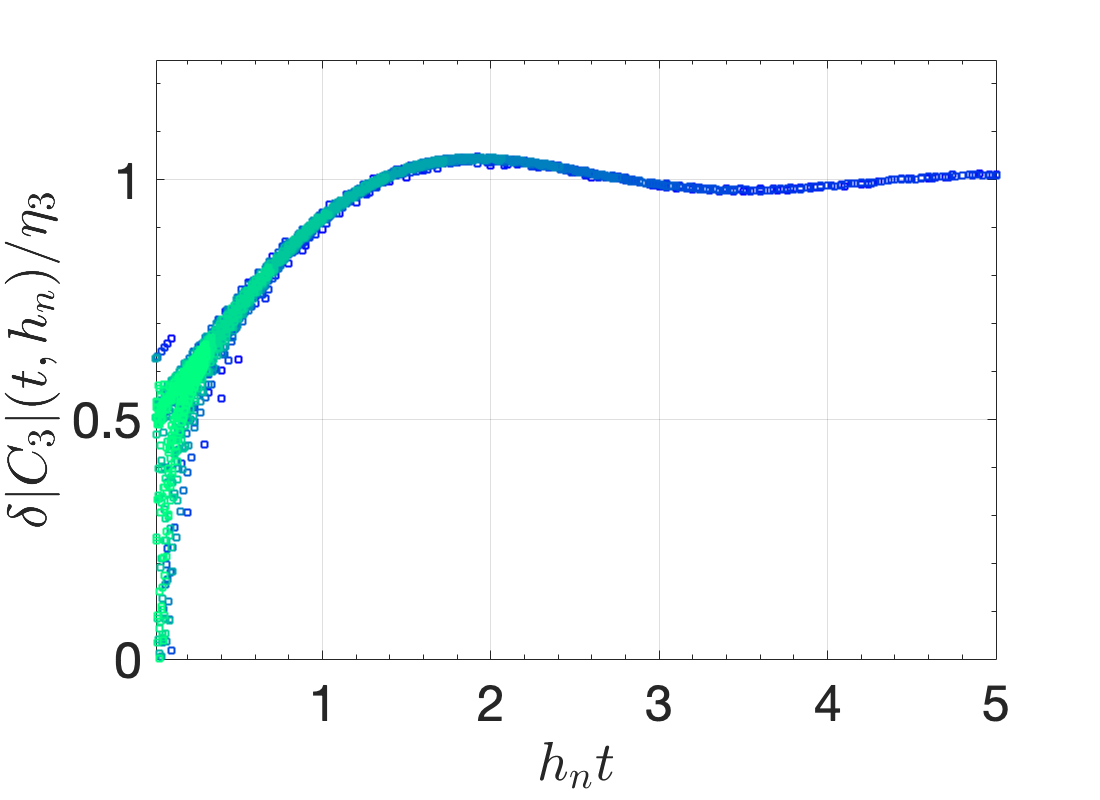}}\hfill
\subfloat{\label{supFig5c}\includegraphics[width=0.33\textwidth]{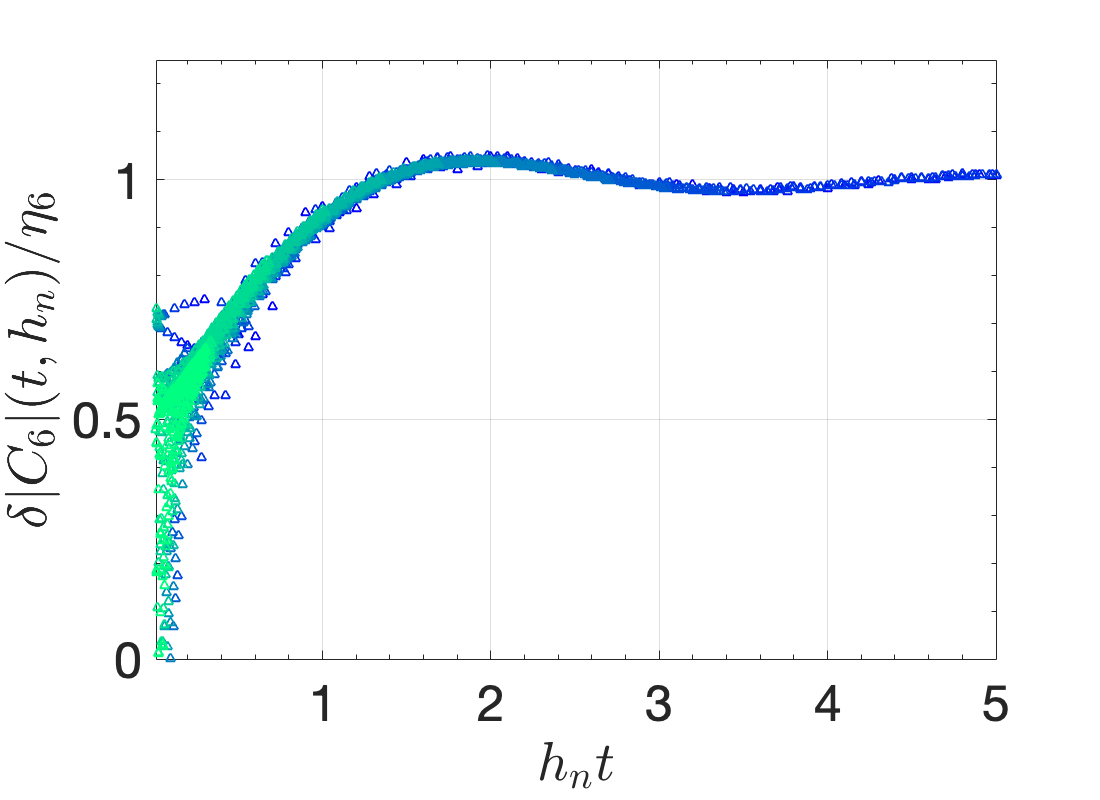}}\hfill
\caption{Scaled critical response $\delta |C_{r}|(t,h_n)/\eta_r$ quenched from $h_i=0$ to $h_n\in [9\times 10^{-4},0.05]$ for the integrable TFIC with a system size of $N=1440$ and for probe sites (a) $r=1$, (b) $r=3$ and (c) $r=6$. The rescaling factor is $\eta_r=C_{1}^{qs} \delta |C_{r}|(t=280,h)/\delta |C_{1}|(t=280,h)$. The data uses the entire time interval $\delta|C_r|(t \geq 0,h_n)/\eta_r$ as opposed to the figures in the main text, and hence shows the early time data points for each $h_n$.} \label{suppFig5}
\end{figure*}

Due to the oscillatory behavior in the early times, c.f.,~$J_0(4t)$, we expect a slight mismatch between the early time numerics and the analytics. This mismatch is demonstrated in Fig.~\ref{suppFig5} for $r=1,3,6$. We observe from numerics that the early time oscillations depend on $r$, suggesting that these oscillations are nonuniversal part of the time evolution.

\section{The dynamical order parameter for $\Delta \neq 0$ and $h_i \neq 0$}

\begin{figure}
\subfloat{\label{supp3}\includegraphics[width=0.45\textwidth]{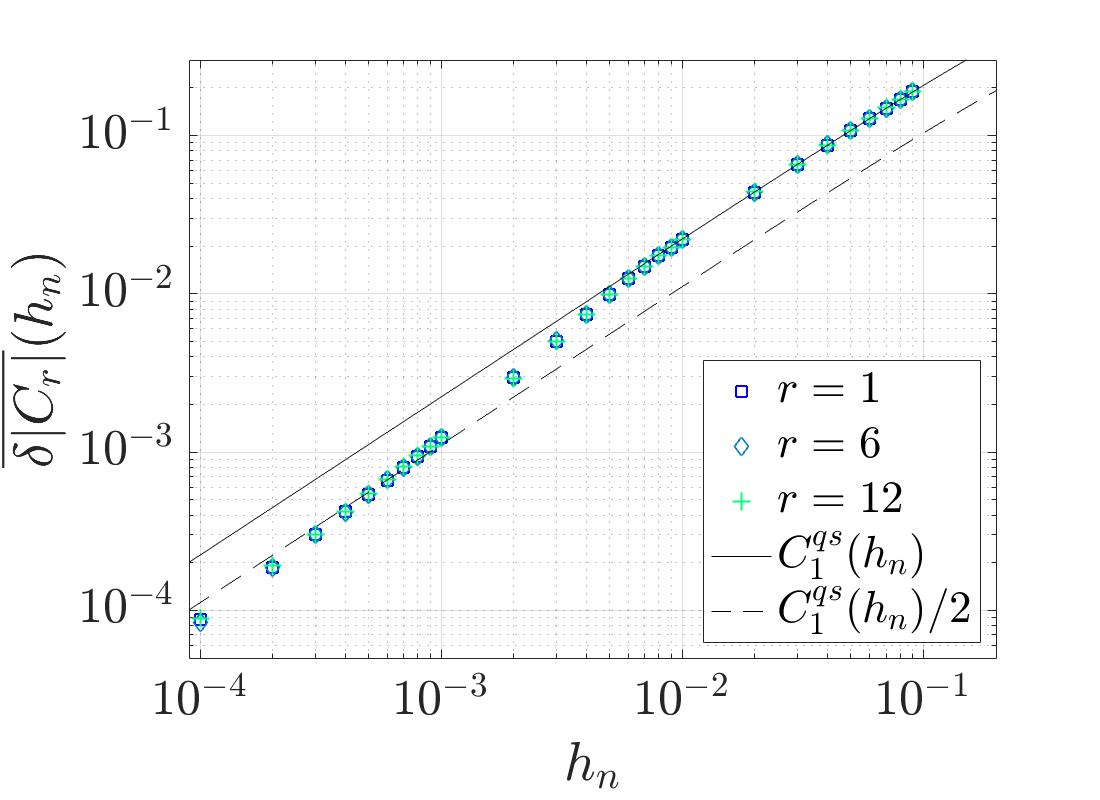}}
\subfloat{\label{supp4}\includegraphics[width=0.45\textwidth]{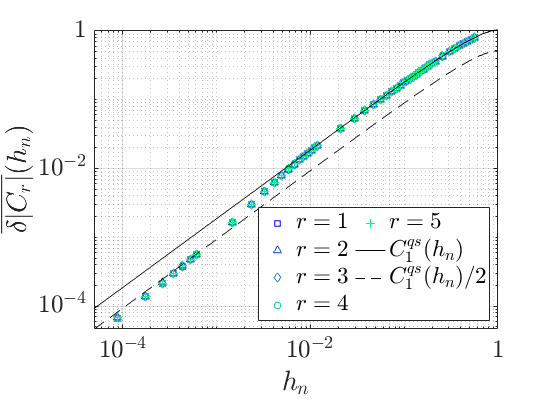}}\hfill
\caption{The dynamical OP for (a) the integrable TFIC with $h_i=0.1$ and cutoffs $t^*=20$ and $t_l=330$ and (b) the near-integrable TFIC with $\Delta=0.1$, $h_i=0$, and cutoffs $t^*=20$ and $t_l=280$. Three different scalings appear corresponding to the different regimes where $t_l$ is in. When $t_l$ is in the decay $(t_l h_n\gg 1)$ or q.s.~$(t_l h_n \ll 1)$~regimes, the data shows linear scalings, described by  $C_{1}^{qs}(h_n)/2$ (dashed-black line) and $C_{1}^{qs}(h_n)$  (solid-black line) of the corresponding parameter set, respectively. When $t_l$ is in the prethermal regime $(h_n t_l\sim 1)$, the scaling diverges from the two linear functions.}
\end{figure}

In this section, we study the dynamical OP Eq.~\eqref{DOP} in the case where we either have a different initial state than a polarized state, $h_i=0.1$ or we introduce weak interactions $\Delta=0.1$. We already showed the presence of a critically prethermal regime in both cases, however the exact form of the function is different than $f_{0,0}(h_n t)$. Then as expected, the corresponding dynamical OP equations for both will be different than Eq.~\eqref{analyticalScaling}. Nevertheless, a nonlinear scaling is still expected to be present due to the existence of prethermal regime. Figs.~\ref{supp3} and~\ref{supp4} show the DOP scaling for these cases. In both, we observe the same function for different probe sites after a rescaling as discussed in the main text. Specifically, we rescale the data using $\eta_r$ and plot $\overline{\delta |C_r|}(t_l,h_n)C^{qs}_1(h_n)/\eta_r$. Here note that $C^{qs}_1(h)=\frac{(1-h^2) (1-2 h_i)}{(1-2 h h_i)}$ generally for $0 \leq h_i \leq h_c$. For the same parameters, the distance to the QCP is $h_n=1-h$. For $\Delta=0.1$ and $h_i=0$, the q.s.~regime expression is given in Ref.~\cite{PRBSub} as $C^{qs}_1(h)=\alpha (h_{dc}^{\nu}-h^{\nu})$ where $\alpha=0.81$, $\nu=1.81$ and $h_{dc}=1.1437\pm0.0001$. In this case, the distance to the DCP is $h_n=(h_{dc}-h)/h_{dc}$.

\section{Independency of the linear-to-linear crossover from temporal cutoffs}

\begin{figure}
\subfloat{\label{supp1}\includegraphics[width=0.45\textwidth]{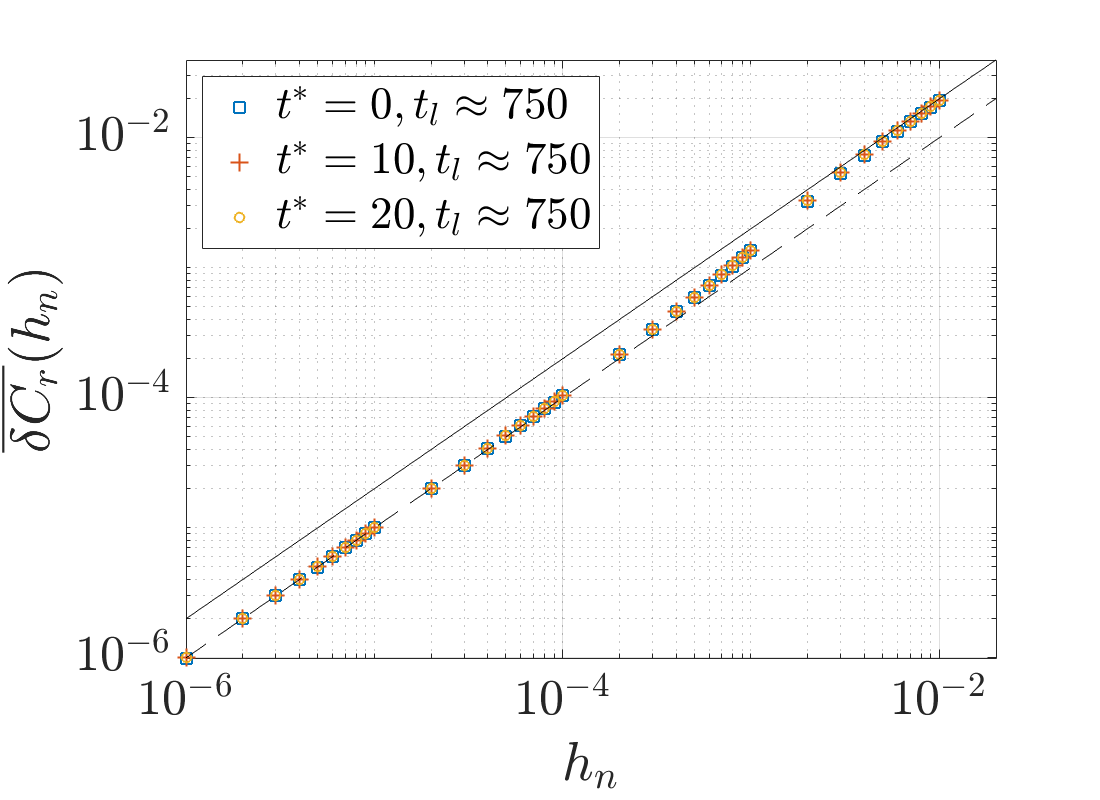}}
\subfloat{\label{supp2}\includegraphics[width=0.45\textwidth]{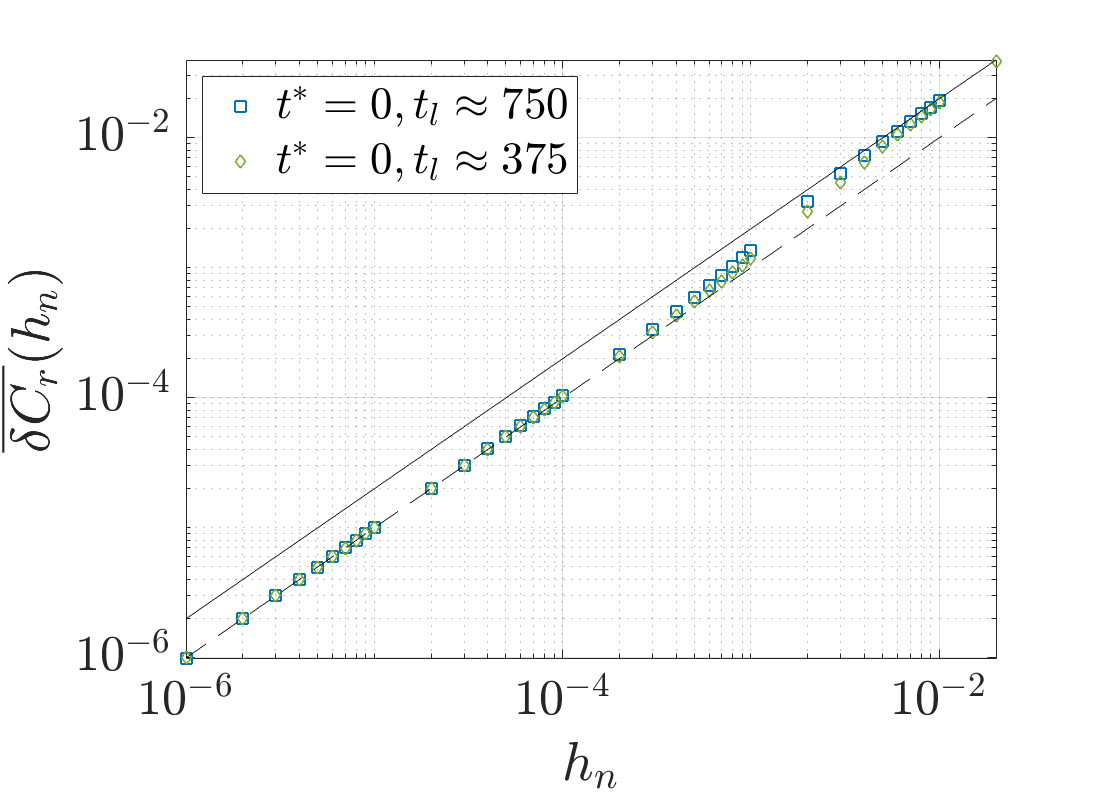}}\hfill
\caption{Scaling of the dynamical OP at the edge calculated from analytical series Eq.~\eqref{series} in the main text, for (a) different short-time cutoff $t^*$ and (b) different long-time cutoff $t_l$. Solid and dashed-black lines are the linear scalings coming up in the q.s.~regime and in the beginning of the prethermal ramp, respectively.}
\end{figure}

We demonstrate the independency of the results from the choice of temporal cutoffs, short-time and long-time cutoffs in Figs.~\ref{supp1} and~\ref{supp2}, respectively. Choosing a larger $t_l$ enlarges the numerical region where $C^{qs}_1(h_n)$ matches, however eventually the scaling becomes nonlinear when $t_l$ traverses through the prethermal ramp. Similarly, the scaling regions do not significantly change when a different short-time cutoff $t^*$ is chosen, so long as $t^* \ll t_{pt}$.

A heuristic way of explaining why a nonlinear scaling should exist is the following: When the temporal cutoff $t_l$ traverses from $h_n t_l > 1$ to $h_n t_l \ll 1$ as $h_n \rightarrow 0$, the scaling has to remain linear up to a change of its nonuniversal coefficient $2h_n \rightarrow h_n$. When $t_{pt}< t_l <t_{qs}$ holds, since the prethermal regime lasts long enough with a ramp, it allows for a nonlinear scaling to emerge. Although there is no analytical expression to our knowledge for the q.s.~regime of the bulk spins, the presence of a universal critically prethermal regime helps for the same scaling to show up for bulk spins, Fig.~2c in the Letter, as well as for a different initial state or for the weakly interacting TFIC (Appendix E).

\section{The analytical form of the rescaling parameter $\eta_r$ for $\Delta=0$ and $h_i=0$}

\begin{figure}
\subfloat{\label{suppFig3a}\includegraphics[width=0.45\textwidth]{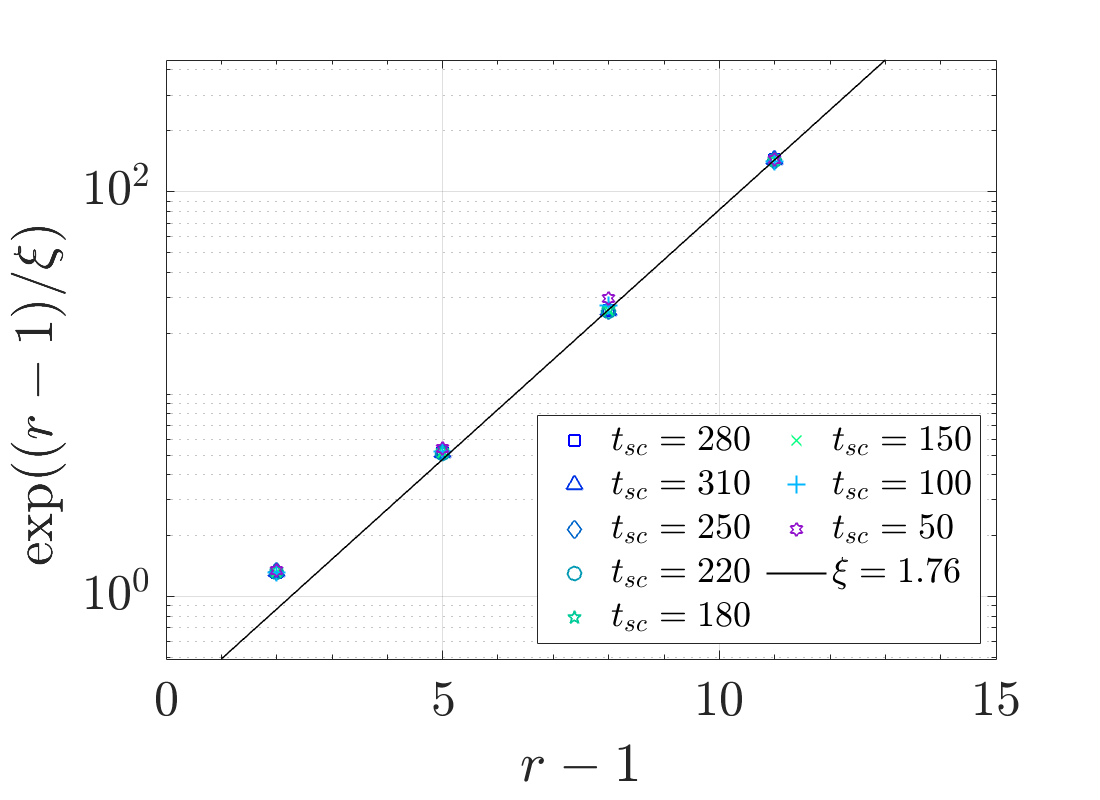}}
\subfloat{\label{suppFig3b}\includegraphics[width=0.45\textwidth]{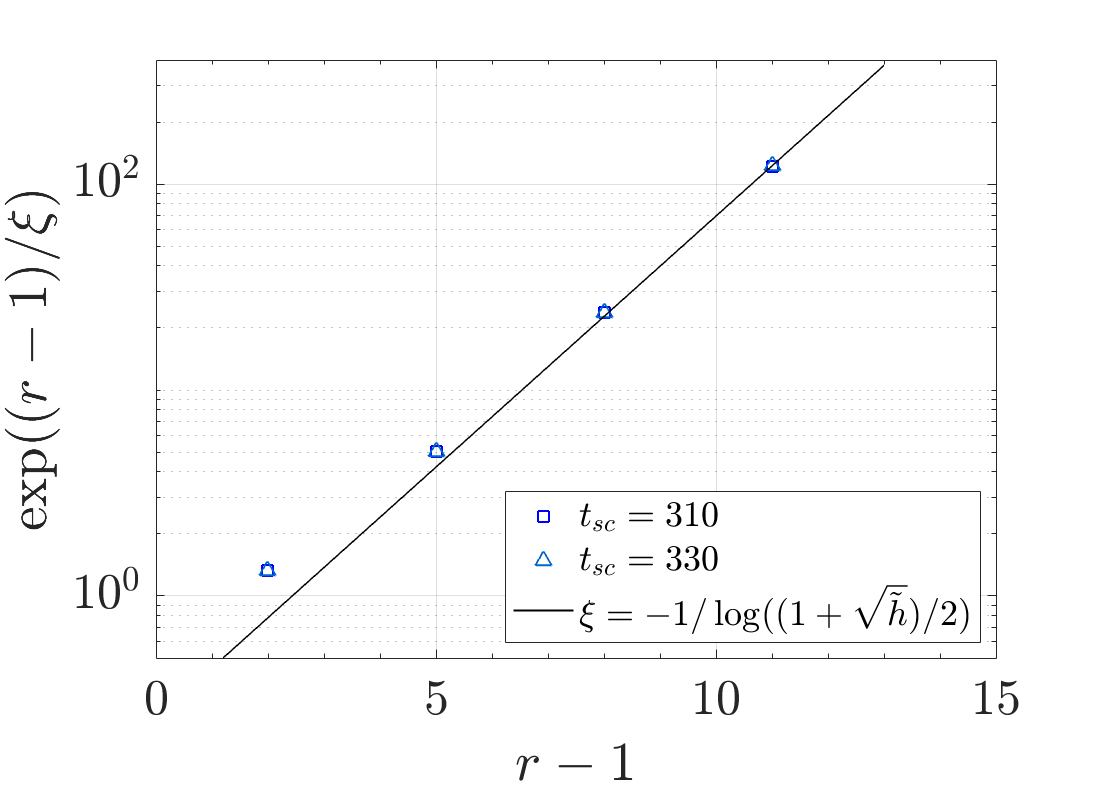}}\hfill
\caption{The exponential scaling that emerges when $\delta C_r(t,h)$ at any $r$ is rescaled to collapse on $\delta C_{1}(t,h)$, (a) at $h=0.995$ and $t_{sc}$ is chosen in the prethermal regime and (b) $h=0.99$ and $t_{sc}$ is chosen in the q.s.~regime. In q.s.~regime analytically predicted $\xi$ is observed when $r \gg 1$ is sufficiently large. In prethermal regime, $\xi$ deviates from this prediction, but still keeps the exponential form for $r \gg 1$.}
\end{figure}

The rescaling parameter $\eta_r$ can be analytically written for $\Delta=0$ and $h_i=0$. This is specifically $\eta_r \approx C_{1}^{qs}(h_n)e^{(r-1)/\xi(h_n)}$. We confirm the exponential form numerically when the scaling time $t_{sc}$ is chosen to be in the q.s.~regime in Fig.~\ref{suppFig3b}. In fact in this case the correlation length follows as $\xi = -1/\log((1+\sqrt{1-h^2})/2)$. When $t_{sc}$ is chosen to be in the prethermal regime, Fig.~\ref{suppFig3a} confirms the exponential form of $\eta_r$ again, however the correlation length diverges from $-1/\log((1+\sqrt{1-h^2})/2)$. Let us note that, changing $h_i$ and $\Delta$ should in principle not change this exponential form of rescaling parameter $\eta_r$.

\end{document}